# Surface Coil Winding Profile for Obtaining a Uniform Magnetic Field inside a Spheroidal Body[1]

*Yavuz Öztürk[a]\*, Alican Aktaş[b] and Bekir Aktaş[b]*

[a] University of Cambridge, Department of Engineering, Electrical Engineering Division, 9 JJ Thomson Avenue, CB3 0FA, Cambridge, UK.

[b] Magnomech Ltd., KOSGEB, Cayirova, 41420, Kocaeli, Türkiye.

\*Corresponding Author: yo268@cam.ac.uk

## Abstract

Uniform magnetic field generation is one of the key issues in many physical applications; such as magnetic resonance imaging, magnetic probs etc. There are many ways to obtain highly homogeneous fields. For instance, ellipsoidal bodies - and in special, spheroids - naturally maintain uniform magnetic field inside their bodies, without any need for shimming. Thanks to their rotational symmetry in addition, spheroids are easier to produce and handle, and therefore deserve more emphasis. However, creating a uniform magnetic field inside a spheroid is only possible via maintaining certain current profiles on its surface. In this paper we have derived *exact* surface coil winding profiles for both prolate and oblate spheroids by reorganizing state of the art derivations in the literature and correcting them whenever necessary.

## 1. Introduction

Ellipsoidal structures have been known to produce uniform magnetic fields inside, since the time of Maxwell. For instance, Marsh [1] and Blewett [2] showed that magnetic field uniformity can be attained by maintaining a constant ampere per turn ratio along the principal axis, that is, by winding a coil of constant pitch along the major axis. For instance, in the case of z being the major axis, this corresponds to a constant z displacement in each complete turn. Thus, the winding function must be a linear function of z. This fact is well used in both theoretical and experimental studies in the literature for the case of spheroidal [1–7] in general, as well as for spherical structures [8–16].

---

[1] Originally this article was meant to be a Comment to Ref [17]. However, the editor of the journal AEÜ is not convinced to publish it because of out of scope concerns.

Živaljevič and Aleksič [17] made a derivation based on Maxwell's equations to show that certain surface currents on a spheroidal structure can generate uniform internal magnetic fields. However, they concluded in an incorrect surface winding profile (surface coil) to produce that desired surface currents. Therefore, we have rederived the winding function by using Maxwell's equations following the similar derivation procedure throughout all steps they followed. And we have noticed that they misinterpreted the surface winding profile and claimed a direct proportionality between the winding profile and the surface current density. Here we show the correct winding profile.

## 2. Derivations

The derivations of the surface current on spheroids (namely prolate or oblate ellipsoids) start with the assumption that electric and magnetic fields take the following forms in the low frequency regime [17]:

$$\boldsymbol{E}(u,v) = E_w(u,v)\widehat{\boldsymbol{w}} \tag{1}$$

$$\boldsymbol{H}(u,v) = H_u(u,v)\,\widehat{\boldsymbol{u}} + H_v(u,v)\widehat{\boldsymbol{v}} \tag{2}$$

where $\widehat{\boldsymbol{u}}, \widehat{\boldsymbol{v}}, \widehat{\boldsymbol{w}}$ are unit spheroidal coordinate vectors. Making use of the absence of charges inside and outside the spheroid, one can employ a scalar magnetic potential which obeys the Laplace equation in prolate (oblate) spheroidal coordinates.

## 2.1. Prolate Spheroidal Case

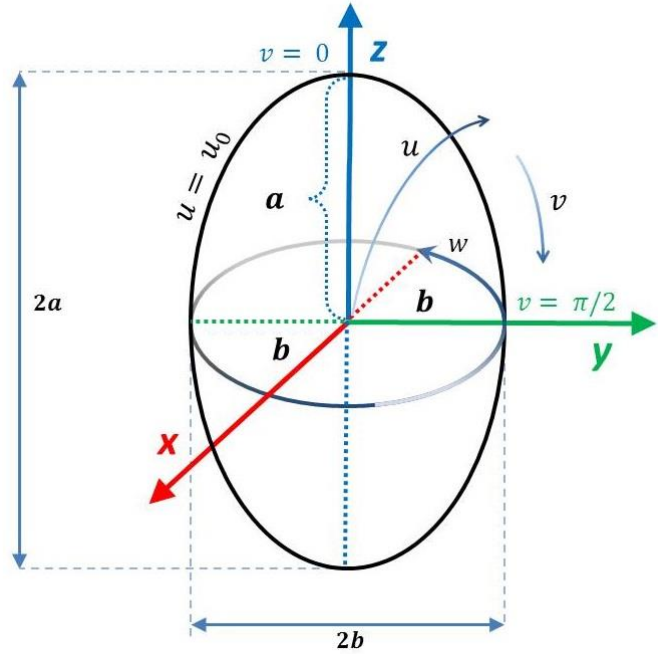

**Figure 1.** Prolate Spheroidal Coordinates

Solving the Laplace equation in prolate spheroidal coordinates (See Figure 1) one can obtain the following expression in Eq. (3) [17].

$$\boldsymbol{H}(u,v) = j_s \left( \frac{\sinh^2 u_o}{2} \ln \frac{\cosh u_0 + 1}{\cosh u_0 - 1} - \cosh u_0 \right) \times (\cot v \sinh u \, \hat{\boldsymbol{u}} - \cosh u \, \hat{\boldsymbol{v}}) \quad (3)$$

for the magnetic field inside. The relations between prolate spheroidal and Cartesian coordinates are given as follows [18]:

$$x = \sqrt{a^2 - b^2} \, \sinh u \, \sin v \, \cos w$$
$$y = \sqrt{a^2 - b^2} \, \sinh u \, \sin v \, \sin w \quad (4)$$
$$z = \sqrt{a^2 - b^2} \, \cosh u \, \cos v$$

where $a$ and $b$ represent the spheroidal dimensions (Figure 1). Here with the use of these relations in Eq. (4), prolate spheroidal unit vectors $\hat{\boldsymbol{u}}, \hat{\boldsymbol{v}}, \hat{\boldsymbol{w}}$ can be expressed in Cartesian unit vectors $\hat{\boldsymbol{x}}, \hat{\boldsymbol{y}}, \hat{\boldsymbol{z}}$ as follows:

$$\hat{\boldsymbol{u}} = \frac{\sqrt{a^2 - b^2}}{h}(\cosh u \sin v \cos w\, \hat{\boldsymbol{x}} + \cosh u \sin v \sin w\, \hat{\boldsymbol{y}} + \sinh u \cos v\, \hat{\boldsymbol{z}})$$

$$\hat{\boldsymbol{v}} = \frac{\sqrt{a^2 - b^2}}{h}(\sinh u \cos v \cos w\, \hat{\boldsymbol{x}} + \sinh u \cos v \sin w\, \hat{\boldsymbol{y}} + \cosh u \sin v\, \hat{\boldsymbol{z}}) \qquad (5)$$

$$\hat{\boldsymbol{w}} = -\sin w\, \hat{\boldsymbol{x}} + \cos w\, \hat{\boldsymbol{y}}$$

We rewrite the result of Živaljevič and Aleksič [17] for the magnetic field:

$$\boldsymbol{H} = -\frac{C_1}{\sqrt{a^2 - b^2}}\hat{\boldsymbol{z}} \qquad (6)$$

for $C_1$ being:

$$C_1 = -j_s \frac{h}{\sin v}\left(\frac{\sinh^2 u_o}{2}\ln\frac{\cosh u_0 + 1}{\cosh u_0 - 1} - \cosh u_0\right) \qquad (7)$$

where $h$ is the Lame coefficient for the prolate spheroidal coordinates defined as

$$h = \sqrt{a^2 - b^2}(\sinh^2 u + \sin^2 v)^{1/2} \qquad (8)$$

Obviously, in order to obtain a constant magnetic field in Eq. (6) aligned along the z axis, $C_1$ has to be a constant. That is, in Eq. (7) $j_s$ must have the form: $j_s \propto \sin v/h$ for this to happen since the terms under the bracket are constants on the spheroidal surface. Introducing a proportionality constant parameter ($j_0$) and using the definition of $h$, the surface current density can be more properly expressed as follows [17]:

$$j_s = \frac{j_0 \sin v}{h} = \frac{j_0 \sin v}{\sqrt{a^2 - b^2}(\sinh^2 u + \sin^2 v)^{1/2}} \qquad (9)$$

Using Eq. (9) Živaljevič and Aleksič supposed the winding function ($N'$) for a prolate spheroidal surface to have the same behavior, namely they expected an incorrect expression for $N'$:

$$N' = \frac{N_0 sinv}{(sinh^2 u + sin^2 v)^{1/2}} \quad (10)$$

where $N_0$ is indicating total number of turns of the coil. In fact their assumption contradicts with the previous studies [1–16] as well. All of these studies emphasized the constant ampere per turn ratio along the principal axis to be obeyed; in order a uniform magnetic field inside to be realized. This wrong conclusion in Eq. (10) can be corrected as follows.

By referring to the time-invariant equation of continuity ($\boldsymbol{\nabla} \cdot \mathbf{j} = 0$) on a closed surface, $\mathbf{j}$ can be represented by curl of a scalar quantity $\phi$ (a differentiable current function) [1],

$$\mathbf{j} = \boldsymbol{\nabla} \times \phi \hat{\boldsymbol{n}} \quad (11)$$

here $\phi \hat{\boldsymbol{n}}$ is a vector off the surface where $\hat{\boldsymbol{n}}$ is the unit normal.

Since $\mathbf{j} = \boldsymbol{\nabla} \times \phi \hat{\boldsymbol{n}} = \boldsymbol{\nabla} \phi \times \hat{\boldsymbol{n}} + \phi \boldsymbol{\nabla} \times \hat{\boldsymbol{n}}$, in which the second term vanishes for a closed surface, the surface current can be written as:

$$\mathbf{j} = \boldsymbol{\nabla} \phi \times \hat{\boldsymbol{n}} \quad (12)$$

Marsh showed that, adopting a linear current function ($\phi = -Kz$, where K is the proportionality constant) with respect to an axis of an arbitrary ellipsoid can provide a uniform magnetic field aligned with the same axis. He consequently showed that for the case of a spheroid, this corresponds to a constant ampere per turn ratio and thus a constant pitch solenoidal coil winding [1].

If we restate Eq. (12) in spheroidal coordinates we obtain:

$$\mathbf{j} = \frac{\hat{\mathbf{u}} \times \hat{\mathbf{n}}}{h_u}\frac{\partial \phi}{\partial u} + \frac{\hat{\mathbf{v}} \times \hat{\mathbf{n}}}{h_v}\frac{\partial \phi}{\partial v} + \frac{\hat{\mathbf{w}} \times \hat{\mathbf{n}}}{h_w}\frac{\partial \phi}{\partial w} \qquad (13)$$

where $h_u = h_v = h$ and $h_w = \sqrt{a^2 - b^2}\, sinhu\, sinv$.

Since $\hat{\mathbf{u}}$ is the surface normal on a spheroidal surface, $\hat{\mathbf{n}}$'s can be replaced by $\hat{\mathbf{u}}$'s. The first term cancels out directly. The last term will also be discarded since $\phi$ has rotational symmetry with respect to $w$. Thus, combining the Eqs. (9) and (13), with the help of the relations in Eqs. (5) the surface current density takes the form retaining only the $\hat{\mathbf{w}}$ components on both sides:

$$j_s = \frac{j_0 sinv}{h} = -\frac{1}{h}\frac{\partial \phi}{\partial v} \qquad (14)$$

where the fact that $\hat{\mathbf{v}} \times \hat{\mathbf{u}} = -\hat{\mathbf{w}}$, is taken into account. Integrating both sides, we have:

$$j_0 \int_{z=0}^{z} sinvdv = -\int_{z=0}^{z} d\phi \qquad (15)$$

$$j_0 \int_{\pi/2}^{v} sinvdv = j_0 cosv = -(\phi(z) - \phi(0)) = -\phi(z) \qquad (16)$$

Here $\phi(0)$ is assumed to be zero as a reference potential. Since we are on a spheroidal surface, $u$ is a constant ($u = u_0$). Inserting $cosv$ from Eq. (4) into Eq.(16), we can write:

$$j_0 \frac{z}{\sqrt{a^2 - b^2}\, coshu_0} = j_0 \frac{z}{a} = -\phi(z) \qquad (17)$$

Defining a new constant $K = j_0/a$ (current density with respect to z axis), Eq. (17) can be rewritten as:

$$\phi = -Kz \qquad (18)$$

which is actually Marsh's linear current function [1]. Thus, starting from Eq. (9) we have arrived at Eq. (18) which rigorously confirms the constant ampere per turn ratio along the principal axis of a spheroid or a sphere. On the other hand, Eq. (17) tells us that $\phi$ is the z integral of the current density, thus is equal to the winding function scaled by the coil current:

$$\phi(z) = IN'(z) \tag{19}$$

Furthermore, it should be remembered that the parameter $j_0$ is the maximum current density which flows at the equator and can be defined in terms of total number of windings $N_0$ as well, by adopting the method used by Haus and Melcher [14]. One should notice that the density of turns along z-axis is constant and equal to $N_0/2a$. Thus, the number of turns in an incremental length is $(N_0/2a)dz$. Since in the prolate spheroidal coordinates the differential may be written as $dz = -a\,sinv\,dv$, the number of turns in the differential length $hdv$ along the periphery is $(N_0/2)(sinv/h)$. $I$ being the coil current, the surface current density is then obtained to be

$$j_s = \frac{N_0 I}{2} \frac{sinv}{h} \tag{20}$$

Comparing Eqs. (14) and (20) we see that the total number of windings $N_0$, coil current $I$ and maximum (equatorial) surface current density $j_0$ are related via:

$$j_0 = \frac{N_0 I}{2} \tag{21}$$

Combining Eqs. (15), (19) and (21) we arrived at the corrected winding function for the spheroidal coordinates which is:

$$N' = \frac{N_0}{2} cosv \tag{22}$$

## 2.1. Oblate Spheroidal Case

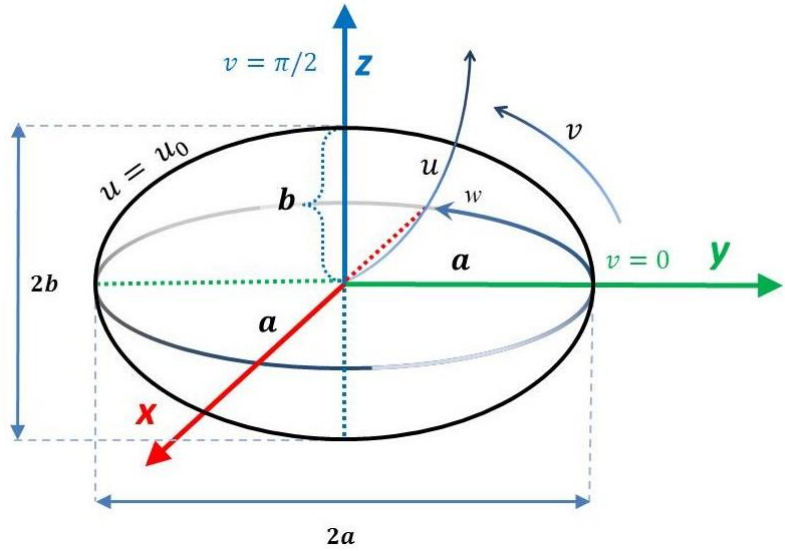

**Figure 2.** Oblate Spheroidal Coordinates

For the oblate spheroidal surface (see Figure 2) current density Živaljević and Aleksić [17] derived:

$$j_s = \frac{j_0 cosv}{h} = \frac{j_0 cosv}{\sqrt{a^2 - b^2}(sinh^2 u + sin^2 v)^{1/2}} \tag{23}$$

Similar to Eq. (10) they concluded the winding function to be:

$$N' = \frac{N_0 cosv}{(sinh^2 u + sin^2 v)^{1/2}} \tag{24}$$

Similarly using Eqs. (12), (13) and (23) we arrive at:

$$j_s = \frac{j_0 cosv}{h} = -\frac{1}{h}\frac{\partial \phi}{\partial v} \tag{25}$$

Integrating both sides again:

$$j_0 \int_0^v cosv\, dv = j_0 sinv = -\phi \tag{26}$$

Relations between oblate spheroidal and Cartesian coordinates are as follows [17,18]:

$$x = c\ coshu\ cosv\ cosw$$
$$y = c\ coshu\ cosv\ sinw \quad (27)$$
$$z = c\ sinhu\ sinv$$

where $a$ and $b$ represent the spheroidal dimensions (Figure 2). Oblate spheroidal unit vectors $\hat{u}, \hat{v}, \hat{w}$ can be expressed in terms of Cartesian unit vectors $\hat{x}, \hat{y}, \hat{z}$ as follows:

$$\hat{u} = \frac{\sqrt{a^2 - b^2}}{h}(\ sinhu\ cosv\ cosw\ \hat{x} + sinhu\ cosv\ sinw\ \hat{y} + coshu\ sinv\ \hat{z}\ )$$
$$\hat{v} = \frac{\sqrt{a^2 - b^2}}{h}(\ coshu\ sinv\ cosw\ \hat{x} + coshu\ sinv\ sinw\ \hat{y} + sinhu\ cosv\ \hat{z}\ ) \quad (28)$$
$$\hat{w} = -sinw\ \hat{x} + cosw\ \hat{y}$$

Using these we arrive at:

$$j_0 \frac{z}{c\ sinhu_0} = -\phi \quad (29)$$

Similar to Eq. (17) and Eq. (18) we have:

$$\phi = -Kz \quad (30)$$

Thus, one can infer that the winding function for the oblate case is:

$$N' = \frac{N_0}{2} sinv \quad (31)$$

Note that $N'$ is proportional to $sinv$ for oblate instead of $cosv$ for prolate case. Considering the definition of coordinates for prolate and oblate cases (Figure 1 and Figure 2 respectively), these two equations (Eqs. (22) and (31)) exactly matches each other for spherical structure.

**As a conclusion:** Starting from the Maxwell's equations for low frequency regime, we have rigorously derived the correct surface winding functions for prolate and oblate spheroidal closed surfaces for producing uniform magnetic fields therein.